\documentclass{article}
\usepackage{amssymb,bbm,amsmath,stmaryrd,subfig,color,mathrsfs,graphicx}
\usepackage{pxfonts}
\usepackage[paperwidth=16cm,paperheight=24cm,textwidth=10cm,textheight=10cm,top=2cm,left=2cm,right=2cm,bottom=2cm]{geometry}
\usepackage{hyperref}
\definecolor{refcolor}{rgb}{0.3,0.3,0.7}
\hypersetup{
    pdftitle={The existence of a critical length scale in regularised friction},
    pdfauthor={David S. Kammer, Vladislav A. Yastrebov, Guillaume Anciaux, Jean-Francois Molinari},
    pdfsubject={Dynamics of frictional contact},
    pdfcreator={LaTeX},
    pdfkeywords={Interface rupture, Regularised friction, Numerical modelling, Finite-Element, Explicit Dynamics},
    pdfnewwindow=true,
    colorlinks=true,
    linkcolor=refcolor,          
    citecolor=refcolor,          
    filecolor=refcolor,      
    urlcolor=refcolor           
}
\usepackage{graphicx}
\usepackage{xspace}
\usepackage{amssymb,amsmath}
\usepackage{marvosym}

\usepackage{xspace}

\newcommand{\ie}{\textit{i.e.}\xspace}
\newcommand{\eg}{\textit{e.g.}\xspace}

\renewcommand{\d}{\mathrm{d}}

\title{The existence of a critical length scale in regularised friction}

\author{David S. Kammer$^a$ \qquad Vladislav A. Yastrebov$^{a,b}$\\ Guillaume Anciaux$^a$ \qquad Jean-Fran\c cois Molinari$^a$\footnote{Corresponding author \href{mailto:jean-francois.molinari@epfl.ch}{jean-francois.molinari@epfl.ch}}}

 \date{\footnotesize$^a${\it Computational Solid Mechanics Laboratory (LSMS, IIC-ENAC, IMX-STI), Ecole Polytechnique
                             F\'ed\'erale de Lausanne (EPFL), B\^at. GC, Station 18, CH 1015 Lausanne, Switzerland}\\
                    $^b${\it Centre des Mat\'eriaux, MINES ParisTech, CNRS UMR 7633, BP 87, F 91003 Evry, France}}

\begin{document}

\maketitle

\begin{abstract}
We study a regularisation of Coulomb's friction law on the propagation of local slip at an interface between a deformable and a rigid solid. This regularisation, which was proposed based on experimental observations, smooths the effect of a sudden jump in the contact pressure over a characteristic length scale. We apply it in numerical simulations in order to analyse its influence on the behaviour of local slip. We first show that mesh convergence in dynamic simulations is achieved without any numerical damping in the bulk and draw a convergence map with respect to the characteristic length of the friction regularisation. By varying this length scale on the example of a given slip event, we observe that there is a critical length below which the friction regularisation does not affect anymore the propagation of the interface rupture. A spectral analysis of the regularisation on a periodic variation of Coulomb's friction is conducted to confirm the existence of this critical length. The results indicate that if the characteristic length of the friction regularisation is smaller than the critical length, a slip event behaves as if it was governed by Coulomb's law. We therefore propose that there is a domain of influence of the friction regularisation depending on its characteristic length and on the frequency content of the local slip event. 
A byproduct of the analysis is related to the existence of a physical length scale characterising a given frictional interface. We establish that the experimental determination of this interface property may be achieved by experimentally monitoring slip pulses whose frequency content is rich enough. 
\end{abstract}


\newpage
\tableofcontents
\newpage

\section{Introduction}
\label{sec:introduction}

Understanding frictional interaction is crucial for studying complex mechanical systems. But it is challenging to gain more insights on friction by studying complex natural (tectonic plates, human joints, etc.) or engineering (gears, bearings, seals, etc.) systems. Thus, experimental research focuses on simplified systems with known properties and measurable behaviour. A wide range of experiments has been conducted on flat interfaces of a given pair of materials in order to analyse the fundamentals of friction~\cite{baumberger:2002,ben-david:2010a,maegawa:2010,rubinstein:2004,xia:2004}. These experiments revealed and analysed several phenomena: different propagation speeds of interface ruptures (from slow to super-shear)~\cite{rubinstein:2004,xia:2004,ben-david:2010a}, crack-like or pulse-like rupture~\cite{coker:2003,baumberger:2002}, and precursor slip events occurring before global sliding~\cite{rubinstein:2004,rubinstein:2007,ben-david:2010a,maegawa:2010}.

Even though such experimental research gives additional opportunities to study phenomena occurring on more complex systems, they bear challenges as well. Accessing information at the interface, where an important part of the studied mechanisms happens, is difficult because the interface is hidden behind the bulk material.
Experimental research often needs to rely on few measurement points distant from the interface, on transparent materials, and on a limited number of experiments. Therefore, numerical simulations are needed to confirm or complement the analysis of these mechanisms. Modelling local slip events at frictional interfaces between dissimilar materials (bi-material interfaces) using mass-spring models~\cite{braun:2009,maegawa:2010,scheibert:2010,amundsen:2012} or the finite-element method~\cite{kammer:2012,diBartolomeo:2012} have shown many similarities with the experiments. However, these numerical models have an inherent problem as explained hereafter.

It was shown~\cite{adams:1995,ranjith:2001} that dynamic sliding of bi-material interfaces under Coulomb's friction law is in many cases unstable, which results in an unbounded increase of displacement oscillations in response to small perturbations at the interface. In real experiments such behaviour has never been observed. 
For the particular case of deformable-rigid interfaces, the stability of Coulomb friction was first studied analytically by~\cite{renardy:1992}, and~\cite{martins:1995}. They showed that sliding of a linear elastic solid on a rigid surface is ill-posed if both the static and kinetic coefficients of friction are greater than one and equal. Moreover, if velocity weakening friction is applied, ill-posedness occurs for smaller friction coefficient as well. 

In numerical simulations the instability due to the bi-material effect results in a lack of mesh convergence~\cite{cochard:2000}. Therefore, most simulations of local slip events need some regularisation to solve this stability problem. Two strategies are known: either the regularisation is applied onto the bulk (\eg, Rayleigh damping or visco-elastic constitutive material) or at the interface (\eg, friction regularisation). In any case, however, it influences the dynamics of local slip events and thus raises questions about the interpretation of numerical results. The solution of interface regularisation is found in the results of experimental work by~\cite{prakash:1993}. They show that the frictional resistance does not change instantaneously to a sudden jump of the normal force, but evolves continuously with time. This observation opposes the Coulomb friction law $F = \mu N$, where the friction force $F$ is proportional to the normal force $N$ with the coefficient of friction $\mu$. Recently, \cite{kilgore:2012} confirmed on a different experimental setup that there is no direct effect on the frictional strength due to a jump in the normal force. A friction law based on these observations introduces a length scale to the definition of the friction force. It was shown that the use of a simplified version of such friction laws renders the bi-material friction problem well-posed and allows to reach mesh convergence~\cite{cochard:2000,ranjith:2001}. This regularisation of friction has since been used widely for earthquake simulations~\cite{rubin:2007,kaneko:2008,brietzke:2009}.

In purpose of avoiding damping in the bulk, we assume here that friction is governed by the Prakash-Clifton law and focus our attention on its effect on the mechanics of slip at frictional interfaces. 
The parameters of this friction regularisation have important implications on the local as well as the global behaviour of friction~\cite{diBartolomeo:2012}, and need therefore to be chosen wisely in order to get meaningful numerical results. However, the choice of appropriate parameters is difficult due to the absence of experimental data for most materials. Furthermore, the characteristic length scale deduced from the experiments of~\cite{prakash:1993} is of the order of micrometres~\cite{cochard:2000}, which requires very fine discretization and heavy computational efforts.

In this work, we study the rupture of interfaces between deformable and rigid solids governed by a regularised friction law. Equal static and kinetic friction coefficients that are smaller than one ensure a well-posed problem even without regularisation~\cite{renardy:1992}. 
Choosing a slip event that is well-posed with and without a friction regularisation avoids a possible distortion of the analysis of the mechanics of regularised friction due to the transition from an ill-posed to a well-posed problem. 
In Section~\ref{sec:influence}, we confirm that mesh-converged solutions are achieved without numerical damping in the bulk for sliding with regularised friction at a deformable-rigid interface. We then depict a convergence map with respect to the characteristic length of the friction regularisation and the discretization of the interface. Using mesh-converged solutions, we show that the friction regularisation has for a given slip event a converging behaviour with respect to the characteristic length of the regularisation\footnote{Notice that throughout this article two different convergences are considered: convergence with respect to the mesh discretization, and with respect to the characteristic length of the friction regularisation.}. The behaviour of a slip event is the same for every characteristic length below the critical length. These observations are confirmed and explained in Section~\ref{sec:filter} by the high-frequency-filter effect of the Prakash-Clifton friction law. The implications of the converging regularisation are analysed in Section~\ref{sec:interpretation} showing that there is a domain of influence for the Prakash-Clifton friction regularisation linked with the characteristic length of the regularisation and the spectral content of the slip event. Outside this domain of influence, the propagation of a slip event under regularised friction is equivalent to the propagation under Coulomb's friction law.

\section{Simulation Setup}
\label{sec:setup}

\begin{figure}[htb!]
  \begin{center}
  \includegraphics[width=0.6\textwidth]{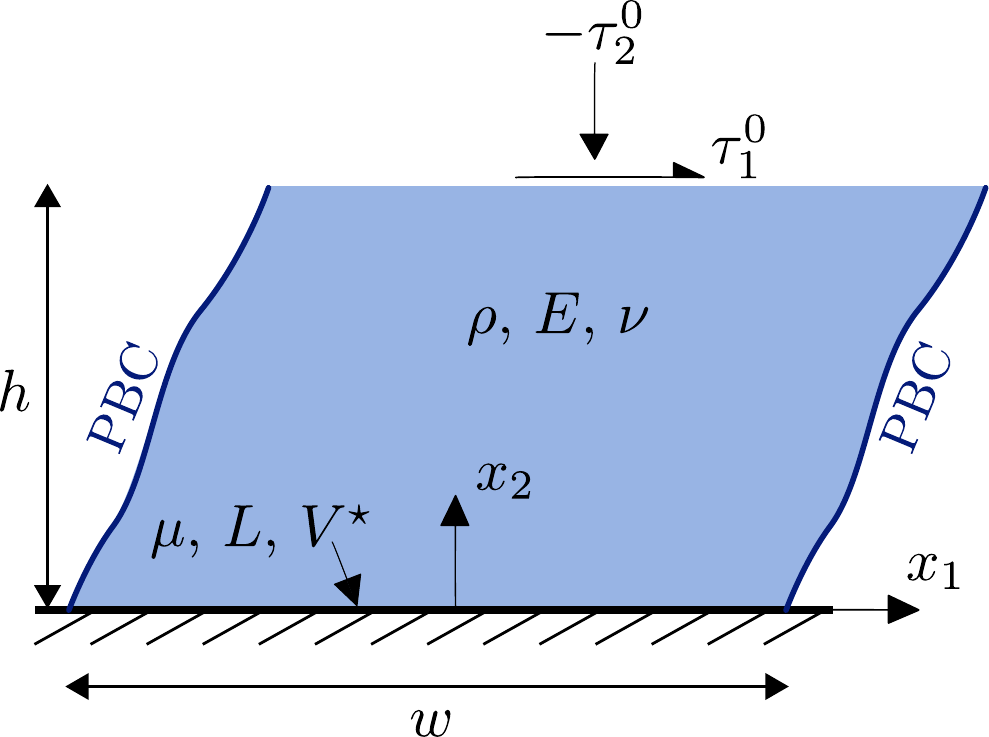}
  \caption{Two-dimensional plane strain model of a frictional interface between an isotropic elastic semi-infinite half-space and a rigid plane. The deformable solid is subjected to a remote static normal and shear load. A rupture, which is nucleated artificially, propagates along the interface until it stops naturally.}
  \label{fig:setup}
  \end{center}
\end{figure}

We study the propagation of a rupture at a frictional interface between a semi-infinite isotropic elastic half-space and a rigid flat surface (Figure~\ref{fig:setup}). This two-dimensional plane strain geometry as well as the material properties are similar to the systems studied by~\cite{andrews:1997} and~\cite{cochard:2000}. We impose in $x_1$ direction Periodic Boundary Conditions (PBC) with replication length $w = 40 \,\textrm{m}$. The height $h = 20 \,\textrm{m}$ ensures that no reflected elastic wave reaches the interface within the time of rupture propagation and does not influence the results of the simulations. The deformable solid is subjected to a remote compressive normal load $-\tau_2^0 = 0.0150 \,\textrm{Pa}$ and a remote shear load $\tau_1^0 = 0.0105 \,\textrm{Pa}$. 

The material properties of the elastic solid are density $\rho = 1 \,\textrm{kg/m}^3$, Young's modulus $E = 2.5 \,\textrm{Pa}$, and Poisson's ratio $\nu = 0.25$. The resulting elastic wave speeds for P (longitudinal) and S (transversal) waves are $c_p = 1.73 \,\textrm{m/s}$ and $c_s = 1.0 \,\textrm{m/s}$, respectively. Friction at the interface is governed by Coulomb's law with equal static and kinetic coefficient of friction $\mu = 0.75$, and is regularised by a simplified~\cite{prakash:1993} law as proposed by~\cite{cochard:2000}
\begin{equation}
\dot \tau_1^s = - \frac{|V| + V^\star}{L}\left[ \tau_1^s - \mu \max (0,-\tau_2) \right] ~,
\label{eq:pc_law}
\end{equation}
where $\tau_1^s$ is the effective frictional strength, $\mu \max(0, -\tau_2)$ is Coulomb friction, $|V|$ is the slip velocity, $V^\star$ is a (positive) reference velocity, and $L$ is the characteristic length of the regularisation. The combination of friction parameters, material properties and imposed loading conditions induces a uniform shear traction at the interface that is at $93.3\%$ of the frictional strength ($\tau_1^0 = 0.933 \mu |\tau_2^0|$). This value enables the propagation of an interface rupture nucleated by an artificial change of normal contact pressure, as it was done in previous studies~\cite{andrews:1997,cochard:2000}. Seeking to make this article self-contained, we describe the nucleation procedure in~\ref{sec:nucleation} as it was already presented in~\cite{cochard:2000}. 
This spatial-temporal nucleation region is of elliptic shape in the $x_1 - t$ plane with $a_{\textrm{\tiny{ell}}}$ and $b_{\textrm{\tiny{ell}}}$ being half the ellipse's minor and major axis, respectively. The parameter $v_{\textrm{\tiny{ell}}}$ inclines the elliptic shape of the nucleation region in the $x_1 - t$, which ensures that the propagation of the interface rupture is oriented in the positive direction of $x_1$ (\ie, the maximum of the artificial change of contact pressure propagates roughly at velocity $v_{\textrm{\tiny{ell}}}$ in the $x_1$ direction). The parameter choice for this study is $a_{\textrm{\tiny{ell}}} = 0.6 \,\textrm{m}$, $b_{\textrm{\tiny{ell}}} = 3.6 \,\textrm{m}$, and $v_{\textrm{\tiny{ell}}} = 0.825 \,\textrm{m/s}$. 
In contrast to previous work, our simulation tool allows for interface opening. In order to avoid such opening and to be consistent with previous studies, we decrease the contact pressure by at most $80 \%$ of its initial value.

The model is based on the finite-element method with an explicit Newmark-$\beta$ integration scheme and a lumped mass matrix. The deformable solid is discretized by regular quadrilateral elements with linear interpolation and four integration points. The mesh is characterised by the node density $n_d$ at the interface, which ranges in the present study from $10$ to $120$ nodes per meter (nd/m). The mesh density is homogeneous in the entire solid in order to avoid spurious wave reflections due to a gradient of mesh density. 

\begin{figure}[htb!]
  \begin{center}
  \includegraphics[width=1\textwidth]{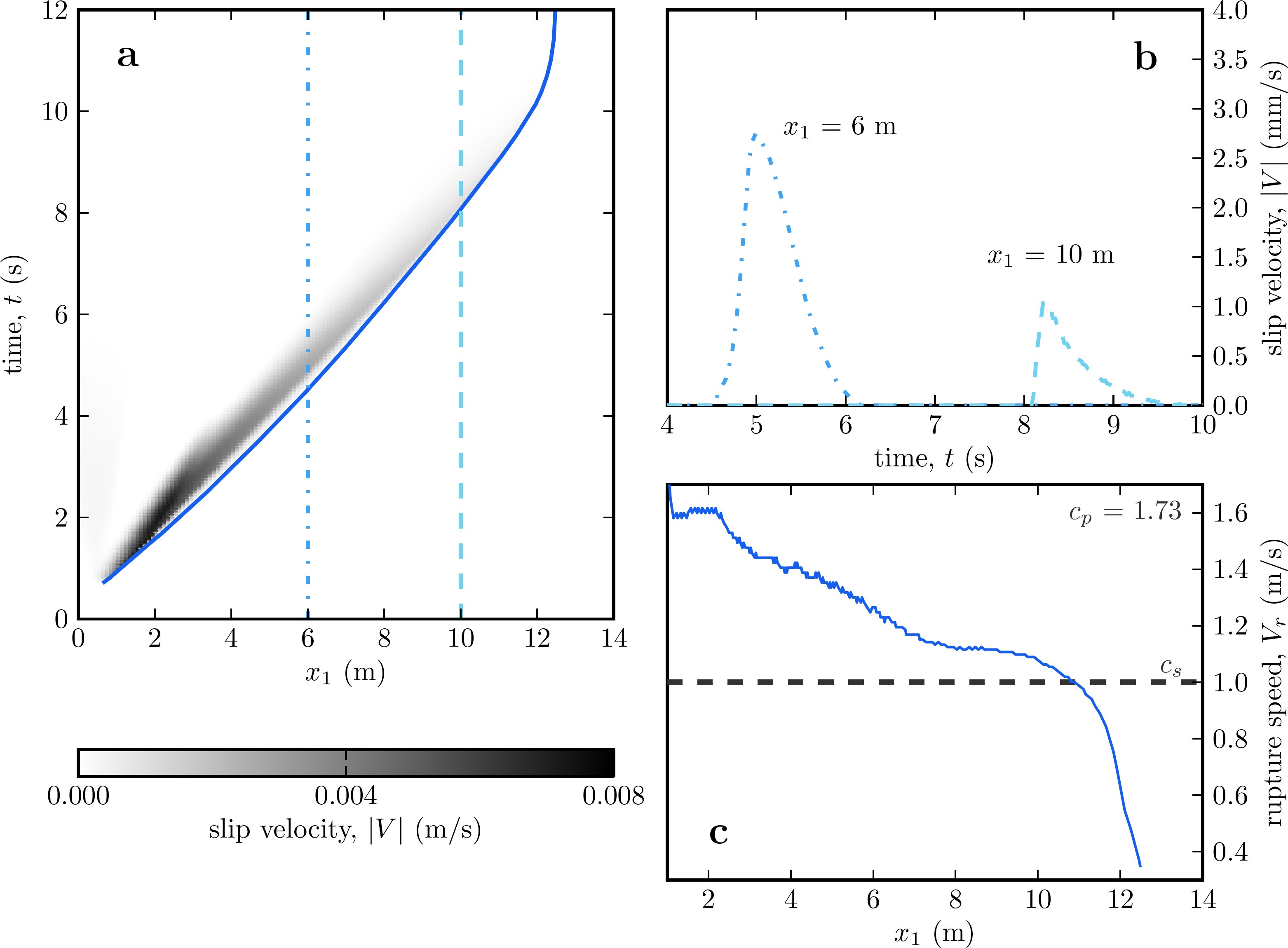}
  \caption{Example of interface rupture with regularisation parameters $V^\star = 10^{-4} \,\textrm{m/s}$ and $L = 5 \cdot 10^{-5} \,\textrm{m}$; and interface mesh density $n_d = 40 \,\textrm{nd/m}$. (a) Slip velocity $|V|$ is shown in the spatial-temporal $x_1 - t$ plane as grey area, where darker colours correspond to higher slip velocities. Vertical lines in (a) indicate positions for which in (b) the evolution of $|V|$ is shown over time. The dark line adjacent to the grey area in (a) marks the front of the interface rupture for which the propagation speed is traced in (c).}
  \label{fig:example}
  \end{center}
\end{figure}

A typical result of a nucleated interface rupture is shown in Figure~\ref{fig:example}. The contact pressure reduction, which triggers the slip event, is located close to the origin. From this point the rupture propagates in the positive direction of $x_1$. The maximal slip velocity is $V_r = 8 \,\textrm{mm/s}$ in the beginning and decreases continuously until the rupture stops at $x_1 \approx 12.5 \,\textrm{m}$. The dark line adjacent to the grey area in Figure~\ref{fig:example}(a) indicates space and time when the slip velocity starts to be non-zero, which we call hereinafter the \emph{slip front}. The light area between this dark line and the dark parts of the grey area shows that the maximal slip velocity does not occur at the front of the interface rupture, which is also notable in Figure~\ref{fig:example}(b). However, the time between the slip front and the maximum of the slip velocity generally reduces with increasing propagation distance.  

The rupture speed $V_r$ shown in Figure~\ref{fig:example}(c) corresponds to the slope of the dark curve in Figure~\ref{fig:example}(a). The rupture propagates with super-shear velocity, $c_s < V_r < c_p$, during almost its entire propagation. It decelerates fast only shortly before the arrest of slip. Even though the slip front in the $x_1 - t$ plane seems to be straight, the rupture speed slows down continuously except for a short distance at $x_1 = 8-10 \,\textrm{m}$, where it remains quasi constant. The decreasing rupture speed along $0 < x_1 < 8 \,\textrm{m}$ is the transition phase before the rupture enters a steady state, which lasts in this simulation for approximately $2 \,\textrm{m}$ until the phase of arrest starts. Other simulations with different regularisation parameters present steady state phases that may span over almost the entire propagation distance. 

In the next section, the slip velocity $|V|$ as well as the rupture speed $V_r$ are analysed for different mesh densities and different values of friction regularisation parameters. We show that the interface ruptures simulated with regularised friction do not only converge with respect to the mesh, but also with respect to the characteristic length of the regularisation.

\section{Influence of Friction Regularisation on Slip}
\label{sec:influence}

The simplified Prakash-Clifton friction regularisation, Eq.~\ref{eq:pc_law}, as proposed by~\cite{cochard:2000}, has two parameters, $L$ and $V^\star$, and depends on the slip velocity $|V|$. In order to simplify the analysis of this regularisation in the present study, we fix $V^\star = 10^{-4} \,\textrm{m/s}$ and consider the variation of the characteristic length $L$. Different values for $V^\star$ result in equivalent observations and conclusions. This simplification leads to a regularisation that still depends on the slip velocity. Other approaches were applied in previous studies~\cite{deDontney:2011a,diBartolomeo:2012}, where the $(|V| + V^\star)/L$ term was replaced by $1/t^\star$. The resulting one-parameter regularisation is similar to our simplification but without dependence on slip velocity.

Before studying the influence of the friction regularisation on the propagation of interface ruptures, we first confirm the mesh-converging quality of the system under consideration and determine a convergence map with respect to the characteristic length $L$. This map enables the choice of an appropriate mesh density for a given $L$ in order to study the effects of the friction regularisation on mesh-converged simulations.

\subsection{Mesh Refinement Analysis}
\label{sec:mesh}

\begin{figure}[htb!]
  \begin{center}
  \includegraphics[width=1\textwidth]{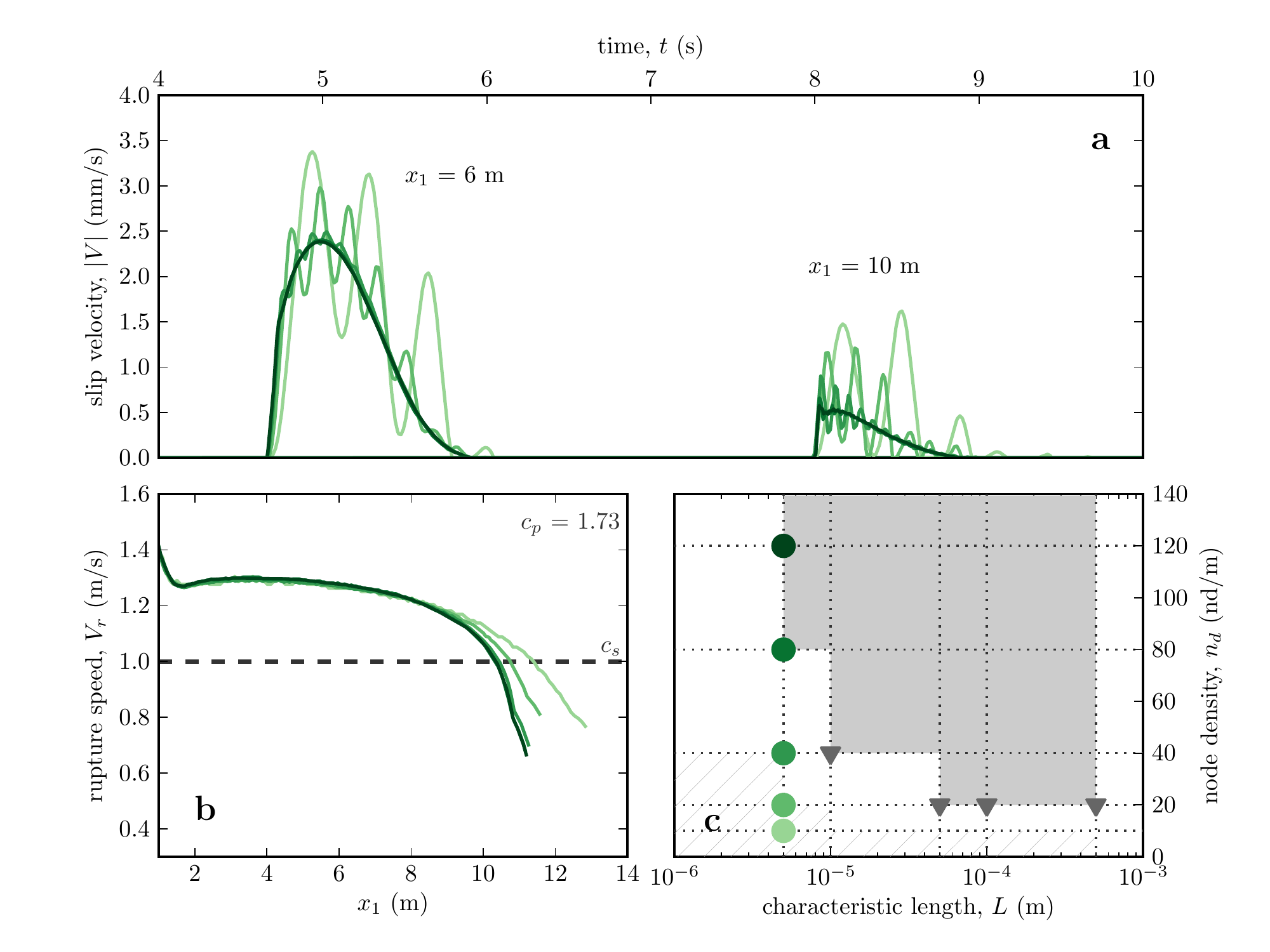}
  \caption{Illustration of mesh convergence for a frictional interface rupture with simplified Prakash-Clifton regularisation $L = 5\cdot10^{-6} \,\textrm{m}$ and $V^\star = 10^{-4} \,\textrm{m/s}$. (a) The evolution of the slip velocity over time is shown at two positions. (b) The rupture speed is depicted with respect to $x_1$. (c) The convergence map in the $L-n_d$ plane indicates the zone of mesh-converged (solid area) and unconverged (hatched area) simulations for $V^\star = 10^{-4} \,\textrm{m/s}$. The filled circles designate the simulations presented in (a) and (b). The triangles mark mesh-converged simulations for different $L$.}
  \label{fig:mesh_conv}
  \end{center}
\end{figure}

The mesh refinement analysis is conducted on meshes with interface node densities $n_d$ ranging from $10 \,\textrm{nd/m}$ to $120 \,\textrm{nd/m}$. The slip velocity and rupture speed for $L=5\cdot 10^{-6} \,\textrm{m}$ and different meshes are shown in Figure~\ref{fig:mesh_conv}, where the colour intensity is chosen accordingly to the mesh density with dark colours being finer meshes. High-frequency oscillations with important amplitudes are present in simulations with coarse meshes. By refining the mesh, the dominant oscillations increase in frequency and decrease in amplitude. 
Mesh convergence is achieved when by refining the discretization the relative error over the slip event's total propagation distance is below $0.5\%$. In addition, we measure the arrival time at $x_1 = 11\,\textrm{m}$. The relative error of this time due to the last mesh refinement is below $0.1\%$ for the case presented in Figure~\ref{fig:mesh_conv}.

In contrast to the slip velocity, the propagation speed of the interface rupture is less affected by the mesh density. The difference can be distinguished only in the arrest-phase, where for the finer mesh the rupture speed decreases significantly faster. A coarser (non-converged) mesh causes the interface rupture to propagate farther and faster during the arrest phase than the converged solution. Compared with the results obtained for $L=5\cdot10^{-5}\,\textrm{m}$ (Fig.~\ref{fig:example}), the rupture speed for this slip event, with $L=5\cdot10^{-6}\,\textrm{m}$, is more steady before arresting and does not present an important decelerating phase before the steady propagation. 

The mesh converging behaviour of this interface rupture is summarised in Figure~\ref{fig:mesh_conv}(c). The filled circles indicate the characteristic length and the node densities of the simulations for which slip velocities and rupture speeds are presented in Figure~\ref{fig:mesh_conv}(a-b). Similar analysis were conducted with different values of $L$. 
The corresponding node densities at the limit of mesh convergence are indicated with triangles.
Simulations with $L=10^{-3}\,\textrm{m}$ present no interface rupture anymore due to the strong regularisation, which prevents initiation of slip for the considered triggering. The solid area marks the zone of mesh-converged solutions, whereas the hatched area shows the zone of non-converged simulations. The limit of mesh-convergence is located in the white area between the two marked zones. The convergence map demonstrates that simulations with smaller characteristic lengths of the friction regularisation need finer meshes for mesh-converged solutions. This observation, which was already noticed without illustration by~\cite{cochard:2000}, is here confirmed and visualised.

A main observation in Figure~\ref{fig:mesh_conv}(a) is that the mesh refinement changes the frequency of the perturbing oscillations. The origin of these oscillations lies in the non-smoothness of the slip velocity in the transition from stick to slip, which excites all ranges of frequencies of the system. A finite discretized solid, however, has only a limited and discrete set of eigenfrequencies. 
Therefore, the energy in the spectral space above the highest representable frequency of the discretization seems to be lumped at this maximal eigenfrequency of the mesh, which is also in the range of the perturbing oscillations. 
Because smaller elements enable the representation of shorter wavelengths, mesh refinement causes higher frequencies of the oscillations. The presence of high-frequency noise is common in this kind of simulations and it is often eliminated by numerical damping in the bulk. Here, we have shown that simulations with regularised friction converge without any bulk damping, which is a necessary condition in order to be able to study the influence of the friction regularisation on the propagation of interface ruptures. 
It is interesting to note that this (damping like) stabilising effect of regularised friction on numerical problems has already been exploited in recent studies on well-posed problems (\eg, frictional interfaces between similar materials)~\cite{deDontney:2012}.

\subsection{Length Scale Convergence}
\label{sec:length}

\begin{figure}[htb!]
  \begin{center}
  \includegraphics[width=1\textwidth]{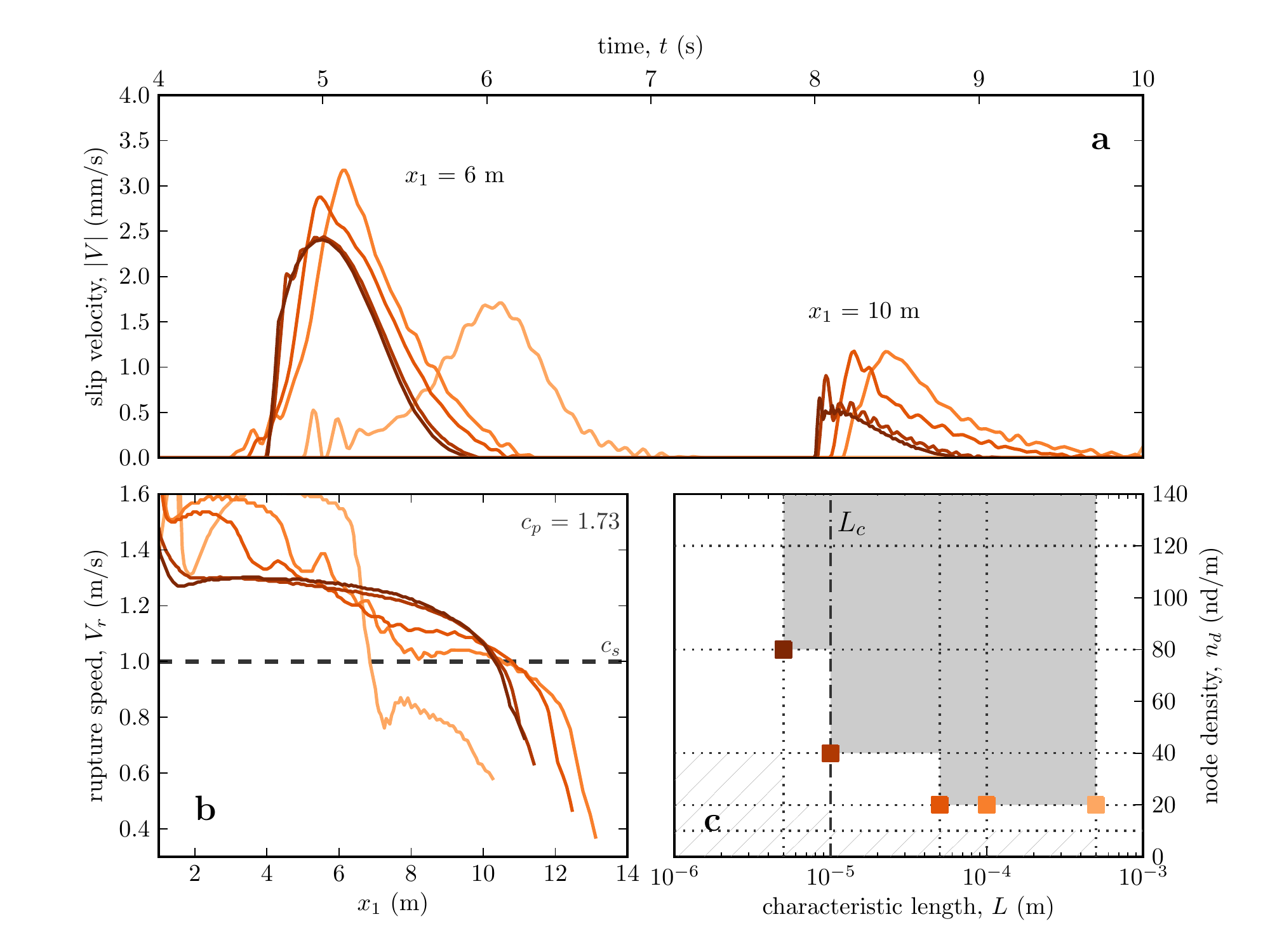}
  \caption{The slip velocity in (a) and the rupture speed in (b) of mesh-converged simulations are shown for different characteristic lengths $L$. (c) The squares indicate the node density $n_d$ and characteristic length $L$ of the simulations presented in (a) and (b). All shown results are in the mesh-converged area of the $L - n_d$ plane. $L_c$ marks the critical characteristic length of the simplified Prakash-Clifton friction law for the presented interface rupture, below which all converged simulations have the same global and local behaviour.}
  \label{fig:L_conv}
  \end{center}
\end{figure}

We now study the influence of the regularisation's length scale $L$ on the slip velocity and rupture speed (see Figure~\ref{fig:L_conv}), similarly as in the mesh convergence analysis. 
The applied meshes are chosen based on the convergence map, as presented in the previous section, at the limit of mesh convergence.
The value of the friction regularisation strongly affects both the slip velocity and the rupture speed, see Fig.~\ref{fig:L_conv}(a) and (b). 
The simulation with the largest characteristic length $L$ (in lightest colour) is the most salient with a considerably smaller slip velocity at $x_1 = 6 \,\textrm{m}$. Further, the slip front reaches $x_1 = 6 \,\textrm{m}$ later than the other simulations despite higher rupture speeds until this position. At a first glance this seems to be incoherent. However, the strong regularisation delays the initiation of the interface rupture, which explains why the slip front is behind the rupture tip of the other simulation despite its higher propagation speed. The interface rupture with the second and third largest characteristic lengths ($L = 10^{-4} \,\textrm{m}$ and $L = 5\cdot 10^{-5} \,\textrm{m}$) have a similar behaviour. Both have at $x_1 = 6 \,\textrm{m}$ a maximal slip velocity of around $3 \,\textrm{mm/s}$ and reach the point at around $4.5 \,\textrm{s}$. Both slip events also present a rupture speed that decelerates from $1.6 \,\textrm{m/s}$ to $1.0 \,\textrm{m/s}$ along almost the complete propagation distance until they arrest abruptly. Similar to the example shown in Figure~\ref{fig:example}, these interface ruptures do not have a phase of steady propagation. 

The two slip events with the smallest $L$ (in darkest colours) obey almost the same propagation behaviour. The slip velocity evolution over time at $x_1 = 6 \,\textrm{m}$ and $x_1 = 10 \,\textrm{m}$ are nearly indiscernible. The most noticeable difference is the rupture speed for $x_1 < 2 \,\textrm{m}$, where a larger characteristic length causes a slightly higher rupture speeds. As already illustrated on the slip event shown in Figure~\ref{fig:mesh_conv}, the propagation speed for interface ruptures with small characteristic lengths presents a steady phase with an almost constant speed during a large part of the propagation distance. Comparing the interface ruptures for all five different characteristic lengths, starting from the largest $L$, we observe a converging behaviour with respect to a decreasing characteristic length $L$. This shows that there is a critical characteristic length $L_c$ below which the choice of $L$ does not influence the propagation of a given slip event. For the interface rupture presented here, the critical length is estimated as $L_c = 10^{-5}\,\textrm{m}$. 
This convergence of the interface rupture with respect to the friction regularisation has not been observed before. We analyse the effect of regularised friction on the spectrum of the slip event in the next section.

\section{Filter Analogy of Frictional Regularisation}
\label{sec:filter}

\begin{figure}[htb!]
  \begin{center}
  \includegraphics[width=1\textwidth]{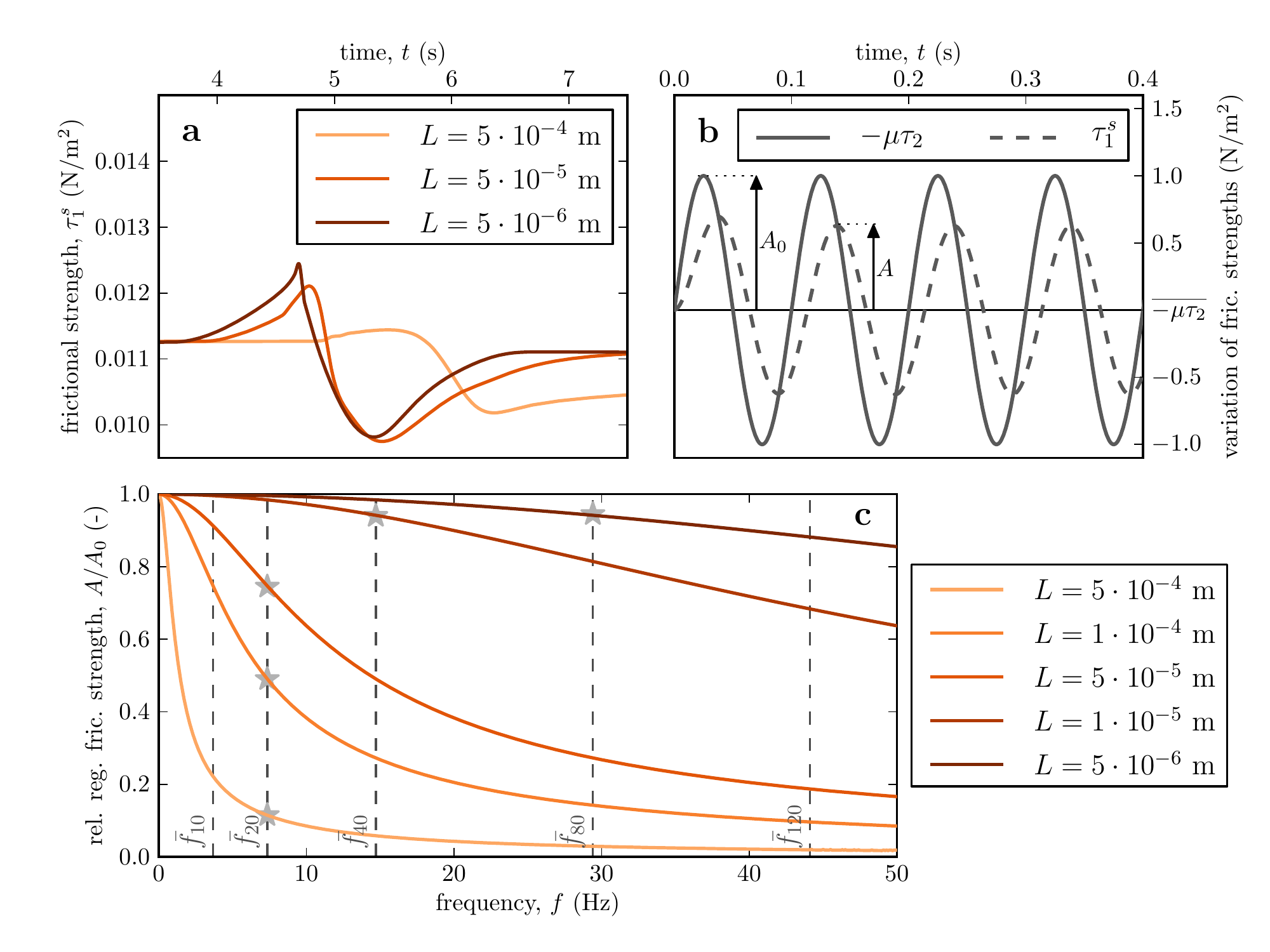}
  \caption{Frequency analysis of the simplified Prakash-Clifton friction law. (a) Variation of the regularised frictional strength ($ \tau_1^s$ in Eq.~\ref{eq:pc_law}) over time at $x_1 = 6 \,\textrm{m}$ for the simulations of triggered interface ruptures as studied in Section~\ref{sec:length}. (b) Spectral analysis of periodic variations in Coulomb friction ($-\mu \tau_2$ in Eq.~\ref{eq:pc_law}) around average value $\overline{-\mu \tau_2}$ with amplitude $A_0$ (solid line) shows that the regularised frictional force $ \tau_1^s$ in Eq.~\ref{eq:pc_law} has a shifted phase, the same frequency and reduced amplitude $A$ (dashed line). (c) The relative regularised frictional strength $A/A_0$  is shown with respect to the frequency of the input signal for friction laws with different characteristic lengths $L$. The dashed vertical lines designated with $\overline{f}_{n_d}$ indicate the highest eigenfrequencies of meshes with node density $n_d$ applied in the finite-element simulations in Section~\ref{sec:influence}. Grey stars mark for each characteristic length the node density needed for mesh-converged solutions.}
  \label{fig:frequency}
  \end{center}
\end{figure}

In Section~\ref{sec:influence}, we observed that the regularisation affects differently the oscillations of various frequencies in the slip velocity $V$. These oscillations are closely related to coupled oscillations of the frictional and normal forces. The oscillations in the frictional force and in the slip velocity vanish with the refinement of the mesh. The evolution of (mesh-converged) regularised frictional strengths for different $L$ at $x_1 = 6 \,\textrm{m}$ are shown in Figure~\ref{fig:frequency}(a). Similarly to the slip velocities, the evolution of the frictional strength during the propagation of the interface rupture is smoother if the regularization $L$ is lager.
Further, the simulation with the smallest characteristic length, which is converged with respect to $L$, has a sharp peak at $t \approx 4.8 \,\textrm{s}$. This feature is related to the fast transition from stick to slip, which is at the origin of the perturbing high frequency oscillations.
The increasing non-smoothness of the transition for smaller regularisations gives a first hint to why finer discretizations are needed in order to achieve mesh-converged solutions for interface ruptures with small $L$.  

We suggest to consider the simplified Prakash-Clifton friction as a (temporal) filter for Coulomb's friction. The physical phenomenon that is regularised is therefore regarded as a signal. The filter receives an input signal, which is Coulomb friction $-\mu \tau_2\left( t \right)$, and creates a filtered (or regularised) output signal, which is the Prakash-Clifton frictional strength $\tau_1^s \left( t \right)$. 
If the filter based on Eq.~\ref{eq:pc_law} receives an input signal with average value $\overline{- \mu \tau_2}$ and a sinusoidal variation over time $t$ of a given frequency $f$ and amplitude $A_0$, the steady-state output signal (for $t >> 1$) is of the same frequency but with a phase offset and a modified amplitude $A$. An example of an input signal (solid line) and the resulting output signal (dashed line) is shown in Figure~\ref{fig:frequency}(b). The analytical solution of the relative regularised frictional strength $A/A_0$ is given by
\begin{equation}
  \frac{A}{A_0} = \frac{1}{\sqrt{1 + \left(\frac{2\pi f}{C} \right)^2}}
\label{eq:aa0}
\end{equation}
with $C = - \left(V^\star + |V|\right) / L < 0$. The complete derivation of Eq.~\ref{eq:aa0} is given in~\ref{sec:aa0}. The filter's influence on the variation of the output signal is independent from the average value $\overline{-\mu \tau_2}$ and from the friction coefficient $\mu$. For physical interpretation, we impose $A_0 < \overline{-\mu \tau_2}$ in order to avoid ambiguity of frictional strength during interface opening.
Concerned by spurious oscillations in the simulations, we neglect the analysis of the phase offset and focus on the influence of the filter on the amplitude of the signal.
The ratio of amplitudes of the output and input signals ($A/A_0$) as a function of the signal's frequency is shown in Figure~\ref{fig:frequency}(c) for filters of different characteristic lengths $L$. The values of $L$ as well as the colour code are the same as for the simulations of interface ruptures presented in Figure~\ref{fig:L_conv}. 
The curves as well as Eq.~\ref{eq:aa0} confirm that the simplified Prakash-Clifton friction law attenuates more the signals of high frequencies. 
It was shown by~\cite{ranjith:2001} that interfacial disturbances present unstable growth for all wavelengths and that the growth rate is inversely proportional to the wavelength. By having a stronger filtering of high frequencies, the friction regularisation compensates for the unstable growth due to interfacial disturbances and solves the problem of the ill-posedness of Coulomb friction at bi-material interfaces.

The signal of a given frequency is more attenuated by a regularisation with larger $L$. This consideration explains the link between $L$ and the node density needed to obtain mesh-converged simulations. The highest eigenfrequencies of the meshes used for the simulations presented in Section~\ref{sec:influence} are indicated by vertical dashed lines in Figure~\ref{fig:frequency}(c). They are labelled with $\bar f_{n_d}$ where $n_d$ specifies the interface node density of the mesh. 
Considering, for instance, the mesh with node density $n_d = 20 \,\textrm{nd/m}$, we see that its highest eigenfrequency is reduced to about $50\%$ for $L = 10^{-4} \,\textrm{m}$. This is sufficient to obtain a mesh-converged solution. A smaller characteristic length, \eg, $L = 10^{-5} \,\textrm{m}$, attenuates a signal of frequency $f=20\,\textrm{Hz}$ only by two percent, which appears to be insufficient for obtaining mesh-converged solutions.
In the previous section, it was shown that a mesh with node density $n_d = 40 \,\textrm{nd/m}$ is needed to obtain mesh convergence for this $L$. The grey stars in Figure~\ref{fig:frequency}(c) indicate for each characteristic length the needed node density for mesh-converged solutions. Two different regimes are observed: 1) simulations with $L \geq 5\cdot10^{-5}\,\textrm{m}$, where mesh-converged solutions are achieved with node density $n_d = 20 \,\textrm{nd/m}$ independently of the attenuation of the highest frequency, and 2) simulations with $L \leq 1\cdot10^{-5}\,\textrm{m}$, where different node densities $n_d$ are needed to obtain mesh-converged solutions. For the latter regime, a mesh-convergence criterion can be defined. The current simulations indicate a need for an attenuation ratio $A/A_0$ of maximal $95\%$ to avoid perturbing oscillations and to obtain mesh-converged solutions. 
On the other hand, slip fronts at interfaces with larger $L$ (regime 1) do not present a particular attenuation ratio that satisfies convergence criterion. The explanation for this absence of criterion lies in the low range of frequencies.
Although the highest frequencies of even coarser meshes seem to be attenuated sufficiently, these frequencies are needed in order to describe accurately the evolution of the frictional strength as shown in Figure~\ref{fig:frequency}(a). Therefore, convergence cannot be achieved with meshes of node densities smaller than $n_d = 20 \,\textrm{nd/m}$ for this particular interface rupture because all needed frequencies are not present with coarser meshes. 

The same argument is valid as explanation of the other convergence: the convergence with respect to the characteristic length $L$ of the Prakash-Clifton friction law. The evolution of the frictional strength over time of this given slip event is mostly composed of low frequencies. 
The mesh convergence regime 2 indicates that these frequencies lie below $\overline{f}_{20}$, which present attenuation ratios that approach one for decreasing characteristic lengths.
Therefore, the influence of the friction regularisation on the evolution of the frictional strength is vanishingly small starting from the critical length $L_c$. 
In our case, we find $L_c = 10^{-5} \,\textrm{m}$, which has a minimal attenuation ratio $A/A_0$ of $0.98\%$ at $\overline{f}_{20}$.
Every $L$ smaller than $L_c$ has negligible influence on the frequencies forming the slip event and every simulation with $L \leq L_c$ obeys the same behaviour, which we call here the convergence with respect to the characteristic length of the friction regularisation. These explanations confirm the observations based on finite-element simulations as reported in Section~\ref{sec:length} and suggest that similar critical length scales should exist for slip events propagating at more general deformable-deformable interfaces.

\section{Physical Interpretation}
\label{sec:interpretation}

\begin{figure}[htb!]
  \begin{center}
  \includegraphics[width=1\textwidth]{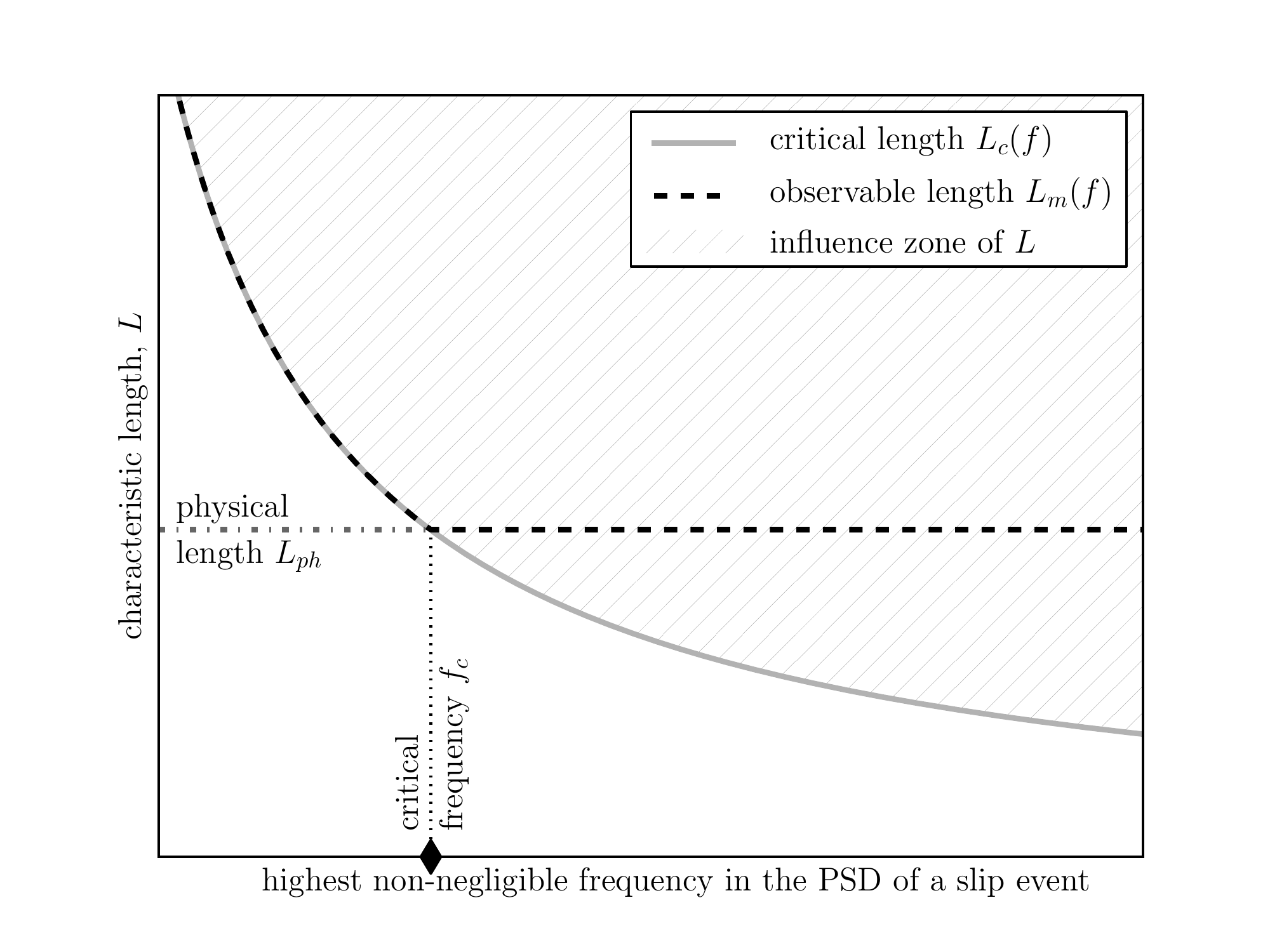}
  \caption{Schematic visualisation of the zone of influence of the Prakash-Clifton friction regularisation. The critical characteristic length $L_c$ decreases for slip events with a power spectral density (PSD) that presents higher non-negligible frequencies. Considering that an interface is not governed by Coulomb's law but actually has a physical length $L_{ph}$, assumed to be independent of the frequency. The intersection of physical and critical lengths defines the critical frequency $f_c$, which separates the slip event's frequencies that propagate under Prakash-Clifton's regularisation ($f>f_c$) and that is governed by Coulomb friction ($f \leq f_c$). Fitting the Prakash-Clifton law to the friction and contact forces of a particular slip event with highest non-negligible frequency $f$ results in the measured length scale $L_m(f)$. If $f \geq f_c$, the measured length scale is the physical property of the interface, whereas if $f < f_c$, the measured length is the critical length, which is higher than $L_{ph}$.}
  \label{fig:L_crit}
  \end{center}
\end{figure}

The critical characteristic length $L_c$ depends on the spectral content of the slip event, which is the result of the nucleation procedure. $L_c$ decreases for slip events reaching higher frequency ranges, as shown schematically in Figure~\ref{fig:L_crit}. The relation between the critical length and the maximal frequency of a slip event is $L_c \propto 1/f$, if a limiting attenuation ratio is assumed in Eq.~\ref{eq:aa0}. The hatched area $L > L_c$ in Figure~\ref{fig:L_crit} indicates the domain, where the parameter $L$ influences the solution. In the white area $L \leq L_c$, the solution is no longer affected by the regularisation. In this zone, the regularised slip propagation is equivalent to the non-regularised propagation (classical Coulomb friction), because the regularisation's influence is vanishingly small.

Let us assume that an interface is not governed by Coulomb friction, but has a physical length scale $L_{ph}$ due to the presence of micro-contacts at the interface. It is important to determine $L_{ph}$ for different interfaces in order to improve our understanding of frictional dynamics. Until today, only few experiments~\cite{prakash:1993,kilgore:2012} have been able to show that friction has a characteristic length $L_{ph}$ with respect to sudden changes in the contact pressure. The value of $L_{ph}$ is estimated~\cite{cochard:2000} to be of the order of microns for the experiments of~\cite{prakash:1993}.
The experimental determination of $L_{ph}$, however, is challenging and our results show that the studied slip events have to be sufficiently rich in high frequencies in order to make the measurement possible.

Considering a particular interface with a given $L_{ph}$, one can distinguish the frequencies that are affected by the characteristic length from the frequencies that are not influenced. This critical frequency is given by $L_c(f_c) = L_{ph}$. Any slip event that has approximately all its frequency content below $f_c$ -- situated to the left of the diamond in Figure~\ref{fig:L_crit} -- is not influenced by $L_{ph}$ and propagates as if the interface was governed by Coulomb's friction law. Therefore, one has to chose carefully the studied slip event and its nucleation procedure in order to determine experimentally $L_{ph}$ of a new interface. The recommended evaluation procedure consists of: 1) experimentally monitor a slip pulse, 2) determine the measured characteristic length $L_m$ by fitting the friction law to the experimental data, 3) compute the measured critical frequency with $L_c(f_m) = L_m$, 4) if the frequency content of the monitored slip event exceeds $f_m$, the procedure was successful and the measured length scale is the physical length of the interface: $L_{ph} = L_m$ and $f_c = f_m$. Otherwise, the studied slip event is not sufficiently rich in high frequencies and the procedure has to be repeated with a sharper slip event, because $L_m = L_c$.

From a numerical point of view, the critical length scale $L_c$ gives in many cases the opportunity of introducing the Prakash-Clifton regularisation in order to solve the issue of perturbing numerical oscillations without influencing the dynamics of the frictional interface rupture. 
Any regularisation with a characteristic length smaller than the critical length does not affect the propagation of the slip event and the observable behaviour corresponds to the propagation under Coulomb's friction law. 
In addition to the effect of removing nuisance oscillations, using the critical characteristic length in numerical simulations minimises the computational efforts because smaller characteristic lengths require finer discretizations.
For instance, if the physical length $L_{ph}$ is smaller than the critical length $L_c$, numerical simulations can be carried out using $L_c$ instead of $L_{ph}$ without losing the physical bases of the results. 
On the other hand, simulations of interface ruptures for which the spectral content above the critical frequency $f > f_c$ is important, does not provide any flexibility in the choice of the regularisation parameters. The characteristic length has to be exactly equal to the physical length $L=L_{ph}$ in order to obtain a correct simulation of the physical behaviour.

\section{Conclusion}
\label{sec:conclusion}

We studied the simplified Prakash-Clifton friction law, which regularises Coulomb's friction with respect to sudden changes in contact pressure. As a test problem we considered the rupture of a planar frictional interface between an elastic solid and a rigid plane. The rupture was triggered artificially such that the slip front propagates in the direction of the movement of the elastic solid and stops after a while. Different stages of the propagation were observed: primary (a transition right after the initiation), steady (the front velocity is almost constant), arresting (the front decelerates and stops). We first confirmed that mesh-converged solutions are achievable in the stable regime (for friction coefficients smaller than one) of the considered problem without any numerical damping in the bulk. We delimited a mesh-convergence map with respect to the characteristic length $L$ of the Prakash-Clifton friction law. This map confirmed that mesh-converged solutions for smaller lengths need finer mesh discretizations. 

In addition to mesh convergence, we discovered a convergence of the solution with respect to the characteristic length $L$. This observation results from the analysis of mesh-converged solutions for different characteristic lengths. Considering a given slip event, a critical characteristic length $L_c$ exists, such that for any $L<L_c$ the propagation behaviour of the interface rupture is the same. To confirm and explain this observation, we analysed the regularisation's effect on a range of temporal frequencies of the frictional strength. The damping of low frequencies, which are the essential part of the slip event, becomes vanishingly small for small characteristic lengths and influences no longer the propagation of the interface rupture. This insight enables the definition of a theoretical domain $L>L_c$ of influence of the Prakash-Clifton friction law with respect to the characteristic length of the regularisation and the frequency content of the slip event. Outside of this domain $L \leq L_c$, the damping of the slip event's frequencies becomes negligible and the interface rupture propagates as if it was governed by Coulomb's friction law despite the presence of the regularisation.

In conclusion, the presented results suggest that the experimental determination of the physical length scale $L_{ph}$ of the Prakash-Clifton friction law requires the temporal power spectrum density of the analysed slip event to contain enough energy in the high-frequency domain. We therefore propose an evaluation procedure that includes a verification of the slip event's frequency content. This is crucial to a successful determination of $L_{ph}$, because if the propagation of slip is fully determined by frequencies below a critical value, the real physical length scale $L_{ph}$ of the Prakash-Clifton friction cannot be measured. The observed length scale instead corresponds to the critical length $L_c$, which may be significantly higher.

\section*{Acknowledgments}
The authors are grateful to M. Radiguet and P. Spijker for fruitful discussions. The research described in this article is supported by the European Research Council (ERCstg UFO-240332).

\appendix
\section{Description of the Nucleation Procedure}
\label{sec:nucleation}

\begin{figure}[htb!]
  \begin{center}
  \includegraphics[width=1\textwidth]{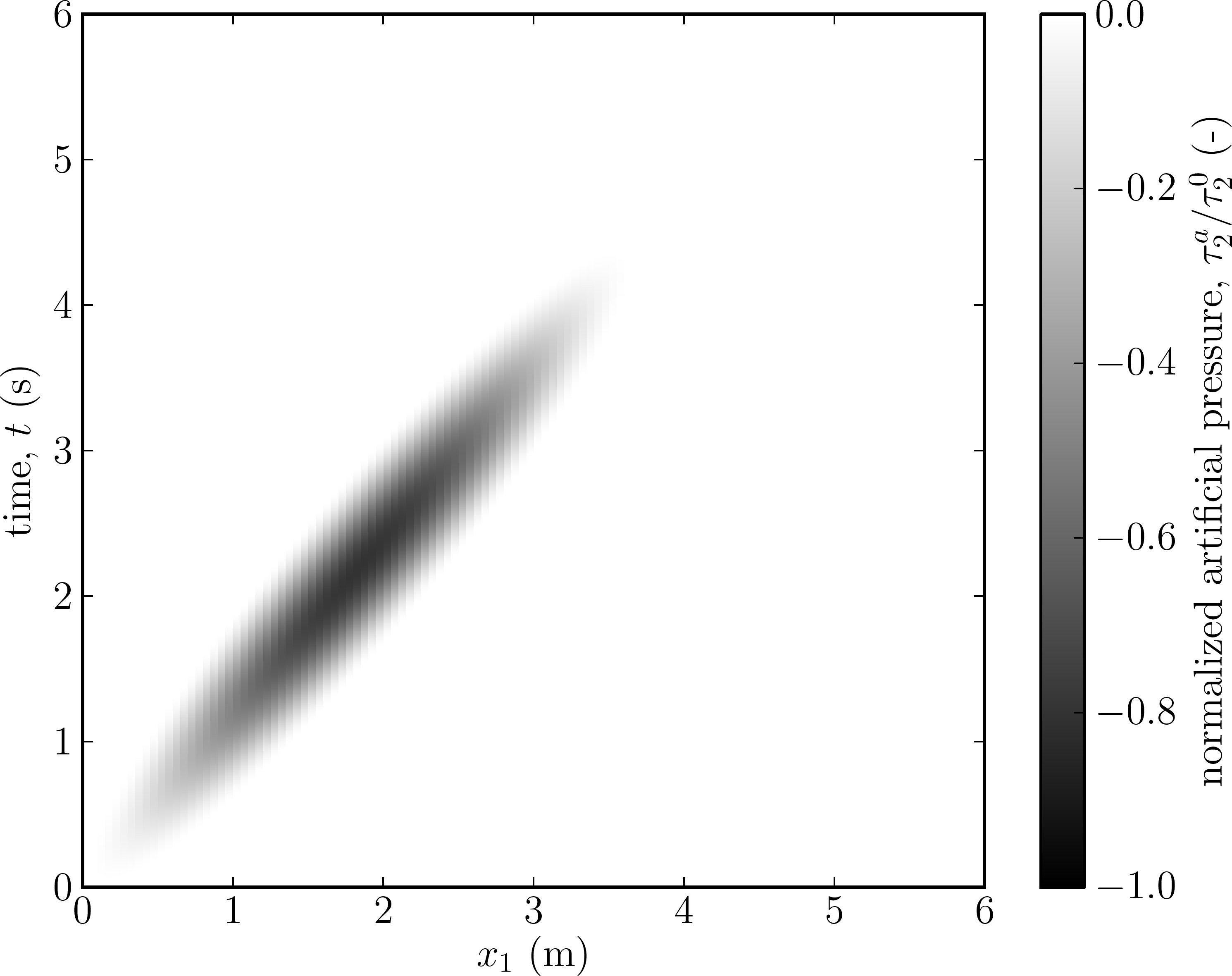}
  \caption{The normalised artificial pressure $\tau_2^a/\tau_2^0$ imposed at the interface is shown for the simulation presented in Figure~\ref{fig:example}. The spatial-temporal region of nucleation is of elliptic shape in the $x_1 - t$ plane.}
  \label{fig:nucleation}
  \end{center}
\end{figure}

The nucleation of the slip events used for this work is here presented following directly the description given in Appendix B of~\cite{cochard:2000}. The procedure consists of an artificial change of the normal contact pressure in a spatial-temporal nucleation region of elliptic shape in the $x_1 - t$ plane. The description is based on following coordinates:
\begin{align}
 \xi &= \left( x_1 - v_{\textrm{\tiny{ell}}} t \right) / a_{\textrm{\tiny{ell}}} \\
 \eta &= \left( x_1 + v_{\textrm{\tiny{ell}}} t \right) / b_{\textrm{\tiny{ell}}} - \eta_0
\label{eq:coord}
\end{align}
with $\eta_0 = \sqrt{a_{\textrm{\tiny{ell}}}^2 + b_{\textrm{\tiny{ell}}}^2}/b_{\textrm{\tiny{ell}}}$. The equation of the ellipse
\begin{equation}
  1 - \xi^2 - \eta^2 = 0
\label{eq:ellipse}
\end{equation}
defines the boundary of the nucleation zone, in which the change of the contact pressure is effected by adding an artificial pressure $\tau_0^a$ (of opposite sign of $\tau_2^0$) to the initial contact pressure $\tau_2^0$. 
The artificial pressure is defined by
\begin{equation}
  \tau_2^a = - \alpha \tau_2^0 \left(1 - \xi^2 - \eta^2\right)^2
\label{eq:pressure}
\end{equation}
where $0<\alpha<1$ is the maximal contact pressure change with respect to $\tau_2^0$.
Outside as well as at the boundary, the contact pressure is equal to the remote, uniform, compressive normal load $\tau_2^0$. Thus, the contact pressure decreases smoothly from $\tau_2^0$ at the boundary down to $\left(1-\alpha\right) \tau_2^0$ at the centre of the ellipse. Here, we choose $\alpha = 0.8$ in order to avoid interface opening, which is not desirable to be consistent with previous studies. The normalised artificial pressure $\tau_2^a/\tau_2^0$ in the nucleation domain is shown in Figure~\ref{fig:nucleation}.

\section{Regularisation: Derivation of the Analytical Solution}
\label{sec:aa0}

In order to improve readability of the derivation, we substitute the physical denotations by the following symbols: Coulomb friction $x = - \mu \tau_2$, regularised frictional strength $y = \tau_1^s$, Prakash-Clifton parameters $C = - \left(V^\star + |V|\right) / L$, and use the angular frequency $\omega = 2 \pi f$.
The simplified \cite{prakash:1993} law as given by Eq.~\ref{eq:pc_law} becomes
\begin{equation}
  \dot y\left(t\right) = C \left( y\left(t\right) - x\left(t\right) \right) ~.
\label{eq:diff}
\end{equation}
Taking a Laplace transform in time, we can rewrite Eq.~\ref{eq:diff} as
\begin{equation}
  \hat y\left(s\right) = \frac{-C}{s-C} \hat x\left(s\right)
\label{eq:laplace}
\end{equation}
with initial condition $y\left(0\right) = 0$. Performing an inverse Laplace transform back to the time domain for $t > 0$, we find
\begin{equation}
  y \left( t \right) = \int_0^t -C \exp{\left(C\tau\right)} \cdot x\left(t - \tau \right) \d \tau ~.
\label{eq:convolution}
\end{equation}
Considering a sinusoidal input signal, we define 
\begin{equation}
  x \left( t \right) = x_0 + A_0 \operatorname{Im} \left( \exp i \omega t \right) ~.
\label{eq:x}
\end{equation}
We then substitute Eq.~\ref{eq:x} into Eq.~\ref{eq:convolution} and simplify to
\begin{equation}
  y \left( t \right) = - x_0 \left[ \exp Ct - 1 \right] + \operatorname{Im} \left[ \frac{-C \exp i \omega t}{C - i \omega} A_0 \left[ \exp \left(C-i\omega\right) t - 1 \right] \right] ~.
\label{eq:y}
\end{equation}
Knowing that $C<0$, we compute the steady-state solution of $y \left(t \right)$ for $t \rightarrow \infty$
\begin{equation}
  y_{\infty} \left( t \right) = x_0 + \frac{A_0}{\sqrt{1 + \left(\frac{\omega}{C}\right)^2}} \sin \left( \omega t + \arctan \frac{\omega}{C} \right) ~.
\label{eq:yinf}
\end{equation}
The steady-state output signal is therefore also a sinusoidal signal around the same average value $x_0$ and with the same (angular) frequency $\omega$. The phase offset is $\phi = \arctan \omega / C$ and the amplitude is 
\begin{equation}
A = \frac{A_0}{\sqrt{1 + \left(\frac{\omega}{C}\right)^2}} ~.
\label{eq:amplitude}
\end{equation}


\end{document}